\documentclass[12pt]{article}
\pdfoutput=1
\usepackage{amsmath}
\usepackage{amssymb} 
\usepackage{bbm} 
\usepackage[small]{caption2} 
\usepackage{fleqn} 
\usepackage{graphicx} 
\usepackage{mathrsfs}   
\usepackage[small,loose]{subfigure}  
\usepackage{cite} 
\usepackage{color}
\usepackage{xspace}
\usepackage{graphicx}
\usepackage{color}

\newcommand{\eVdist}{\kern-0.06em}

\newcommand{\gev}{\:\text{Ge\eVdist V}}
\newcommand{\tev}{\:\text{Te\eVdist V}}

\newcommand{\Z}[1]{\ensuremath{\mathbbm{Z}_{#1}}} 

\newcommand{\be}{\begin{equation}}
\newcommand{\ee}{\end{equation}}
\newcommand{\bea}{\begin{eqnarray}}
\newcommand{\eea}{\end{eqnarray}}

\newcommand{\SARAH}{{\tt SARAH}\xspace}
\newcommand{\SPheno}{{\tt SPheno}\xspace}

\allowdisplaybreaks[1]

\begin{document}

\begin{titlepage}

\vspace*{-3.0cm}
\begin{flushright}
OUTP-12-17P\\
Bonn-TH-2012-20
\end{flushright}

\begin{center}
{\Large\bf
  Enhanced $h\rightarrow \gamma \gamma$ rate in MSSM singlet extensions
}

\vspace{1cm}

\textbf{
Kai Schmidt-Hoberg$^a$,
Florian Staub$^b$
}
\\[5mm]
\textit{$^a$\small
Rudolf Peierls Centre for Theoretical Physics, University of Oxford,\\
1 Keble Road, Oxford OX1 3NP, UK
}
\\[5mm]
\textit{$^b$\small
Bethe Center for Theoretical Physics \& Physikalisches Institut der 
Universit\"at Bonn, \\
Nu{\ss}allee 12, 53115 Bonn, Germany
}
\end{center}

\vspace{1cm}

\begin{abstract}
 We study the di-photon rate in Higgs decays within singlet extensions of the supersymmetric standard model. In particular we point out that light charginos as well as a light charged Higgs 
 can significantly contribute to the corresponding partial decay width, allowing for an explanation of the experimental indication whithin a natural supersymmetric model.
 This is in contrast to the `light stau scenario' proposed within the framework of the MSSM which requires a large amount of electroweak fine tuning.
 \end{abstract}

\end{titlepage}

\section{Introduction}

Recently both ATLAS~\cite{ATLAS:2012} and CMS~\cite{CMS:2012} have presented evidence for a new bosonic state with mass $m \sim 125 \gev$.
While the data is consistent with the expectation from a standard model (SM) Higgs, both experiments see indications for an excess in the 
diphoton channel\footnote{see however \cite{Plehn:2012iz}}, while the diboson decays into $WW^*$ and $ZZ^*$ seem to be in accord with the SM expectation.
In particular the enhanced di-photon rate has attracted much attention recently, see e.g.~\cite{Ellwanger:2011aa,Barger:2012ky,Carena:2012xa,Alves:2012ez,Bonne:2012im,Bellazzini:2012mh,Buckley:2012em,An:2012vp,Cohen:2012wg,Alves:2012yp,Joglekar:2012hb,Haba:2012zt,Almeida:2012bq,Benbrik:2012rm}.
Within the minimal supersymmetric standard model (MSSM) two possibilities to increase the rate $pp \rightarrow h \rightarrow \gamma \gamma$ have been discussed: Higgs mixing effects can have an important impact on the diphoton rate, in particular via a suppression of the coupling to down-type fermions, which leads to an effective increase of the $h \rightarrow \gamma \gamma$ channel~\cite{Barger:2012ky}. However, this would also lead to an increase in the branching fractions into $WW^*$ and $ZZ^*$, which does not seem reflected in the data. The second possibility is to increase the loop induced Higgs photon coupling via some light charged states running in the loop. In the MSSM the only viable possibility was found to be a very light and strongly mixed stau \cite{Carena:2011aa,Carena:2012gp}, whereas the contributions from the charged Higgs and charginos turn out to be negligible. While an interesting proposal, the light stau scenario requires very large $\mu \cdot \tan\beta$, which implies very large 
corresponding electroweak fine-tuning. 

The electroweak fine-tuning can be substantially alleviated in singlet extensions of the MSSM, in part due to the new quartic Higgs coupling, giving an additional tree-level mass contribution to the light Higgs.
The GNMSSM~\cite{Ross:2011xv} based on a discrete $R$ symmetry, $\Z{4}^R$ or $\Z{8}^R$ \cite{Lee:2010gv, Lee:2011dya}, was found to be particularly interesting in this context~\cite{Ross:2012nr}.
The aim of this letter is to point out that in singlet extensions of the MSSM the coupling of the CP even neutral light Higgs to charginos
as well as to the charged Higgs can be strongly enhanced, leading to a sizeable increase in the $h \rightarrow \gamma \gamma$ rate.
This allows for an explanation of the enhanced rate within a much less fine-tuned scenario.
We find that this enhancement is rather sensitive to the value of the trilinear singlet-Higgs coupling and is significant only
for values of $\lambda$ which become non-perturbative below the GUT scale, corresponding to the well known scenario of $\lambda$SUSY \cite{Barbieri:2006bg}, which has been argued to be favoured by fine-tuning considerations~\cite{Hall:2011aa}. It has also been argued recently that the unification of gauge couplings might even improve in such a setup~\cite{Hardy:2012ef}. In this letter we don't assume any particular UV completion but rather study the GNMSSM at the electroweak scale. 

This letter is organised as follows: In the next section we briefly review the partial decay width of the Higgs boson $h$ into two photons with a particular focus on the charged Higgs
and chargino contributions. To this end we will perform a simple analysis using tree-level masses to highlight the main features.
In Section~\ref{sec:spheno} we will then perform a complete numerical analysis at one-loop including the leading two-loop contributions in the Higgs sector.
Section~\ref{sec:conclusions} contains our summary and conclusions.

\section{The $h \rightarrow\gamma\gamma$ partial decay width}
To start our discussion let us briefly review the partial decay width of the Higgs boson $h$ into two photons within the MSSM and its singlet extensions.
It can be written as (see e.g.~\cite{Djouadi:2005gj})
\begin{align}
\label{eq:decaywidth}
\Gamma_{h \rightarrow\gamma\gamma} 
& = \frac{G_\mu\alpha^2 m_{h}^3}
{128\sqrt{2}\pi^3} \bigg| \sum_f N_c Q_f^2 g_{h ff} A_{1/2}^{h} 
(\tau_f) + g_{h WW} A_1^{h} (\tau_W) \nonumber \\ 
& \hspace*{1.6cm} + \frac{m_W^2 g_{h H^+ H^-} }{2c_W^2 
m_{H^\pm}^2} A_0^h(\tau_{H^\pm}) \nonumber \\ 
& \hspace*{1.6cm} +  \sum_{\chi_i^\pm} \frac{2 m_W}{ m_{\chi_i^\pm}} g_{h \chi_i^+ 
\chi_i^-} A_{1/2}^{h} (\tau_{\chi_i^\pm}) + 
\sum_{\tilde f_i} \frac{ g_{h \tilde f_i \tilde f_i} }{ 
m_{\tilde{f}_i}^2} \, N_c Q_{\tilde f_i}^2 A_0^{h} (\tau_{ {\tilde f}_i}) \bigg|^2 \,,
\end{align}
corresponding to the contributions from charged SM fermions, W bosons, charged Higgs and charginos as well as the contributions of charged sleptons and squarks.
In the SM the largest contribution is given by the $W$-loop, while the top-loop leads to a small reduction of the decay rate.
In this work we will focus in particular on the contributions from charged Higgs and charginos running in the loop. From Eq.~(\ref{eq:decaywidth}) it can be seen that
the contribution from the charginos is suppressed by $m_{\chi_i^\pm}$, while the contribution from the charged Higgs is suppressed by $m_{H^\pm}^2$.
The amplitudes $A_i$ at lowest order for the 
spin--1, spin--$\frac{1}{2}$  and spin--0 particle contributions are given by \cite{Djouadi:2005gj}
\begin{align}
A_{1/2}^h(\tau)  &=  2 [\tau +(\tau -1)f(\tau)]\, \tau^{-2}  \nonumber \\   
A_1^h(\tau)      &=  - [2\tau^2 +3\tau+3(2\tau -1)f(\tau)]\, \tau^{-2}\nonumber \\
A_{0}^h(\tau)    &=  - [\tau -f(\tau)]\, \tau^{-2} 
\end{align}
with $\tau_i=m^2_{h}/4m^2_i$ and $f(\tau)=\arcsin^2 \sqrt{\tau}$ for $\tau \le 1$.

\begin{figure}[hbt]
\begin{minipage}{\linewidth}
 \includegraphics[width=0.456\linewidth]{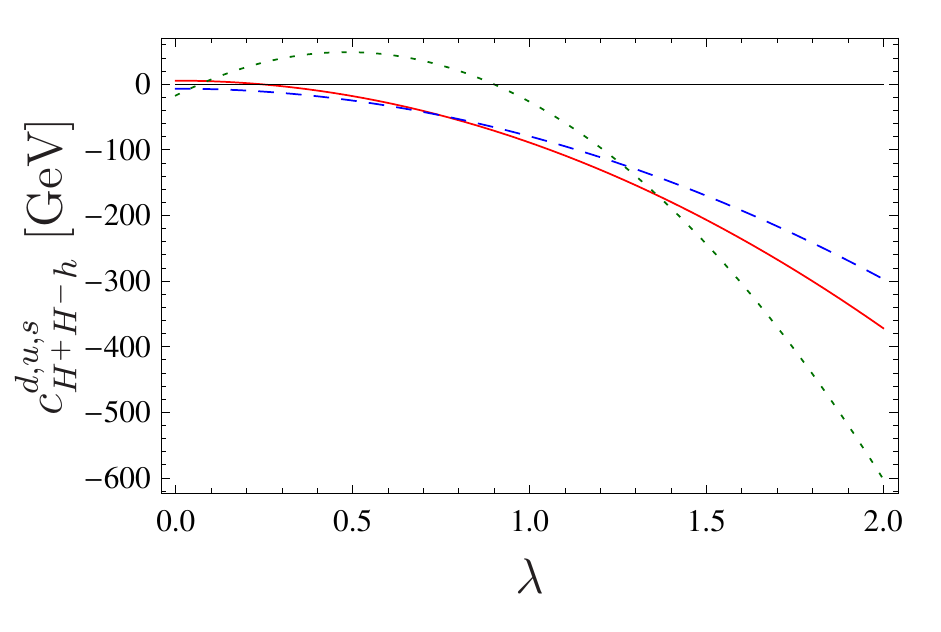} \hfill
\includegraphics[width=0.45\linewidth]{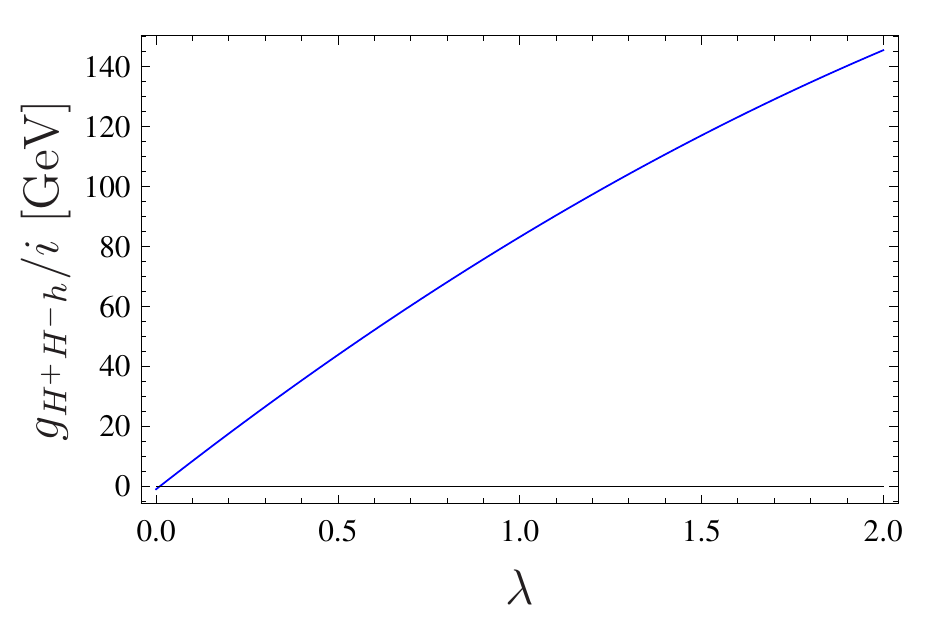} 
\end{minipage}
\caption{Left panel: size of the different contributions to the coupling $g_{h H^+H^-}$ proportional to the down-fraction (red), up-fraction (blue) and singlet-fraction (green) as a function of $\lambda$. Right panel: The overall coupling $g_{h H^+H^-}$ as a function of $\lambda$. Note, we have kept all rotation matrices involved in the vertex constant to show only the dependences on $\lambda$. The values of the additional parameters in both cases have been $\tan\beta=1.3$, $\kappa = 1.0$, $\kappa A_\kappa = -500$~GeV, $\lambda A_\lambda = -25$~GeV, $v_s = 282$~GeV. We calculated the Higgs mixing matrices for $\lambda = 1.0$ and obtained for the scalar Higgs $Z^h_{i} \simeq (-0.23, -0.91, 0.35)$.}
\label{fig:HpLoop}
\end{figure}

In the following we will analyse the partial Higgs decay width into two photons within a generalised version of the NMSSM, which has a superpotential of the form
\begin{eqnarray}
\label{eq:superpotential}
 \mathcal{W} &=& \mathcal{W}_\text{Yukawa}  + \frac{1}{3}\kappa S^3+
(\mu + \lambda S) H_u H_d + \xi S+ \frac{1}{2} \mu_s S^2  \label{gen}\\
&\equiv& \mathcal{W}_\text{NMSSM}+
\mu H_u H_d + \xi S+ \frac{1}{2} \mu_s S^2  \label{gen2} 
\end{eqnarray}
where $\mathcal{W}_\text{Yukawa}$ is the MSSM superpotential generating the SM Yukawa couplings and $ \mathcal{W}_\text{NMSSM}$ corresponds to the normal NMSSM with a $\Z{3}$ symmetry. 
We use the freedom to shift  the vev $v_s$ to set the linear term in $S$ in the superpotential to zero, $\xi=0$. 
Such a superpotential can arise from an underlying $\Z{4}^R$ or $\Z{8}^R$ symmetry as discussed in \cite{Lee:2011dya}.
The corresponding general soft SUSY breaking  terms associated with the Higgs and singlet sectors are given by
\begin{align}
 V_\text{soft} 
   &=  m_s^2 |s|^2 + m_{h_u}^2 |h_u|^2+ m_{h_d}^2 |h_d|^2 \nonumber \\
   &+ \left(b\mu \, h_u h_d + \lambda A_\lambda s h_u h_d + \frac{1}{3}\kappa A_\kappa s^3 + \frac{1}{2} b_s s^2  + \xi_s s + h.c.\right) \;.
\label{soft}
\end{align}
We give the mass matrices of the neutral and charged Higgs particles as well as of the neutralino and chargino in Appendix~\ref{app:massmatrices}. 
For more details on the model, see \cite{Ross:2012nr}. The MSSM or NMSSM limits can easily be obtained by setting the appropriate parameters to zero.

\subsection{Enhancing the di-photon rate with a charged Higgs}

\begin{figure}[hbt]
\begin{minipage}{\linewidth}
 \includegraphics[width=0.45\linewidth]{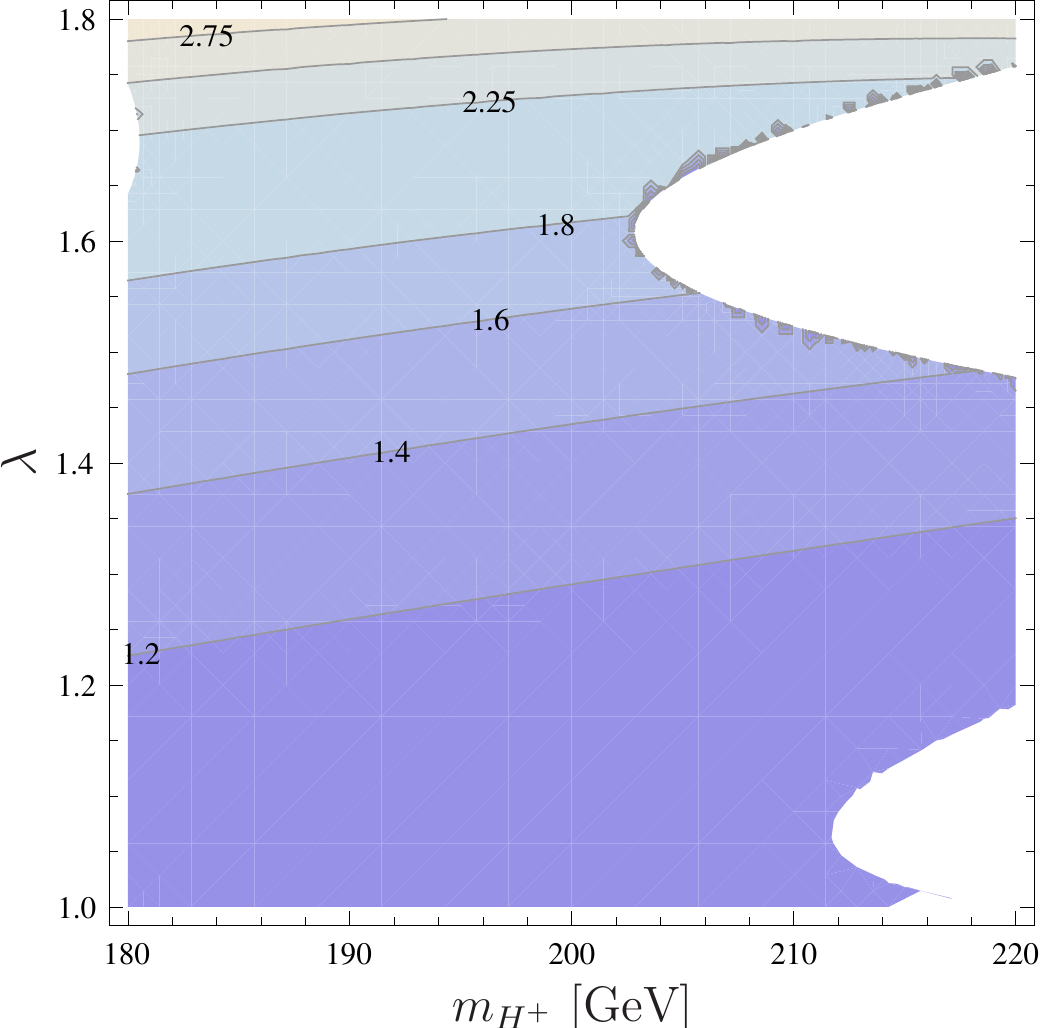} \hfill
\includegraphics[width=0.45\linewidth]{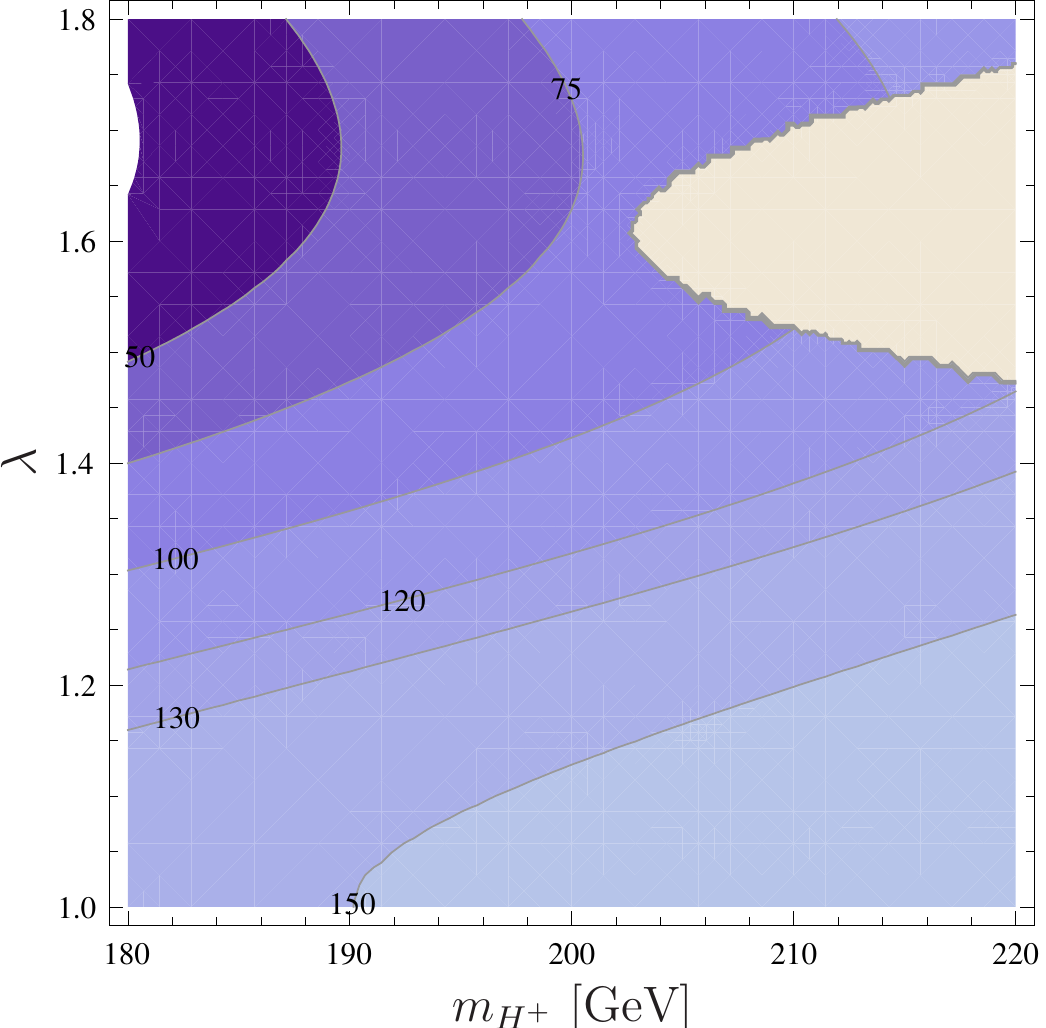}  
\end{minipage}
\caption{Approximate analysis at tree level in the $(m_{H^\pm},\lambda)$ plane. Left panel: partial di-photon width from the charged Higgs normalized to the SM (top and $W$) contributions.  Right panel: \emph{tree-level} mass of the light doublet Higgs.  The input parameters have been   $\tan\beta = 1.3$, $\kappa=1.1$,  $\mu_\text{eff} = 220$~GeV and $\kappa A_{\kappa} = -500$~GeV. }
\label{fig:HpTree}
\end{figure}

The coupling $g_{h H^+H^-}$ in the GNMSSM is given by (taking all parameters real for simplicity)
\begin{align}
g_{h H^+H^-} = & \,
\frac{i}{4} \bigg\{ v \cos\beta \Big[2 (\lambda^2-g_2^2) + (g_1^2 + g_2^2 - 2 \lambda^2) \cos 2\beta \Big] Z^h_1 \nonumber \\
&-   v \sin\beta \Big[2 (g_2^2 - \lambda^2) + (g_1^2 + g_2^2 -  2 \lambda^2) \cos 2\beta \Big]  Z^h_2 \nonumber \\
&-  4 \lambda \Big[v_s \lambda + \sqrt{2} \mu +    (\tfrac{1}{\sqrt{2}} (A_\lambda + \mu_s)  +   v_s \kappa) \sin 2\beta\Big] Z^h_3 \bigg\}
\end{align}
where $Z^h_i$ denote the entries of the Higgs mixing matrix in the $(h_d, h_u,s)$-basis.

\begin{figure}[hbt]
\begin{minipage}{\linewidth}
 \includegraphics[width=0.456\linewidth]{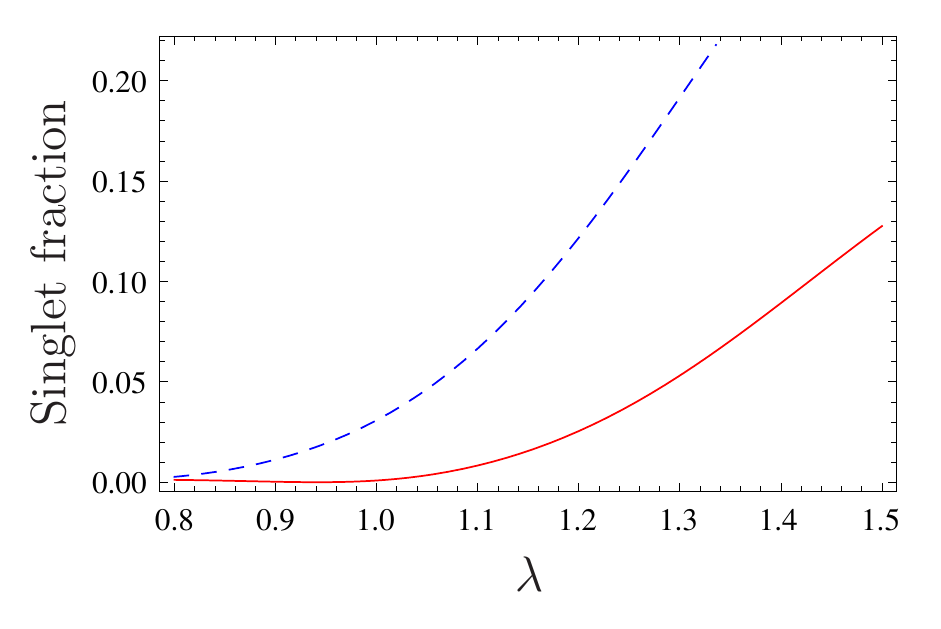} \hfill
\includegraphics[width=0.45\linewidth]{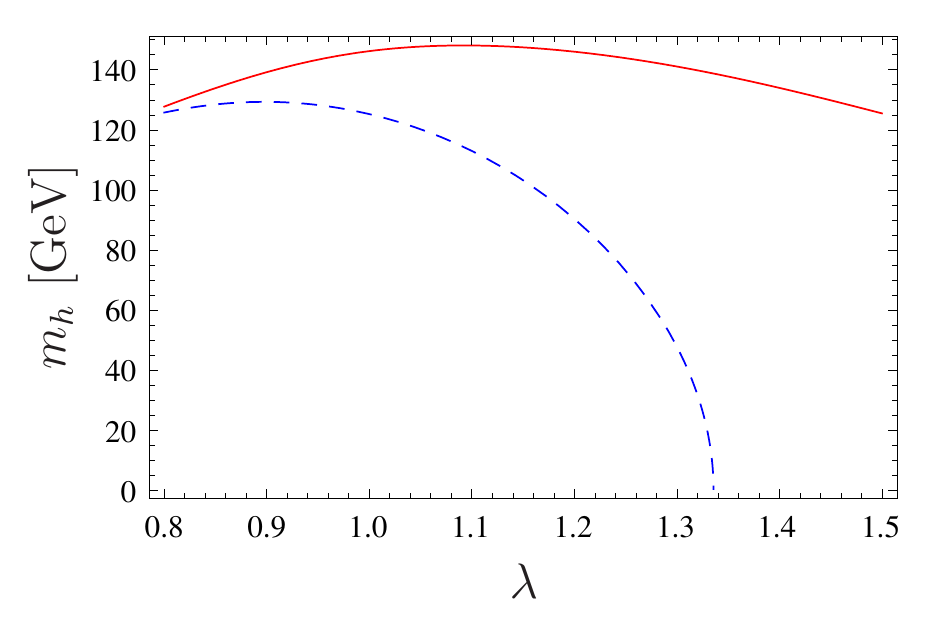} 
\end{minipage}
\caption{Singlet fraction (left) and tree level mass (right) of the light Higgs doublet as a function of $\lambda$. The blue dashed line shows the NMSSM case ($\mu = 0$), while for the red line $\mu = -100$~GeV was used. The other parameters are $\tan\beta = 1.5$, $\kappa = 1.2$, $v_s = 240$~GeV, $m_{H^\pm} = 165$~GeV.  }
\label{fig:mu}
\end{figure}

In the MSSM limit the charged Higgs often reduces the effective coupling of the Higgs to two photons because of negative interference with the others contributions: the sign of the interaction is fixed by the electroweak parameters ($\tan\beta, g_1, g_2, M_Z$) and the Higgs mixing angle.
In the (G)NMSSM, new contributions proportional to $\lambda^2$ and other singlet parameters arise. The terms proportional to the additional singlet parameters couple to the singlet fraction of the light Higgs. Since we are interested in a mostly SM-like Higgs with only a subleading singlet fraction, the new terms proportional to $\lambda^2$ will be most important.
Note that the enhancement due to $\lambda$ is present even if there is no doublet singlet mixing at all.
The coupling $g_{h H^+H^-}$ as a function of $\lambda$ is shown in Figure~\ref{fig:HpLoop}. It can be seen that  $g_{h H^+H^-}$ can be significantly enhanced, leading to a correspondingly enhanced partial di-photon width. This is shown in Figure~\ref{fig:HpTree} as a function of $\lambda$ and $m_{H^\pm}$.
To use the tree-level mass of the charged Higgs directly as input, we solved for $A_\lambda$,
\begin{equation}
A_\lambda = \frac{1}{4 \sqrt{2} \lambda v_s}\left(-4 v_s (\sqrt{2} \mu_s + v_s \kappa)\lambda - 8 b\mu + (4 m_{H^\pm}^2 - v^2 (g_2^2 - 2\lambda^2))\sin(2\beta)\right) \; .
\end{equation}
While $A_\lambda$ is often chosen such that in the NMSSM the doublet-singlet mixing in the CP even Higgs sector is reduced, this freedom is lost by this approach. However, it turns out that the singlet fraction can be kept small by utilising the additional GNMSSM parameters, e.g.\ by choosing moderate finite values of $\mu$ as shown in Figure~\ref{fig:mu}. 

\subsection{Enhancing the di-photon rate with charginos}
\begin{figure}[hbt]
\begin{minipage}{\linewidth}
\centering
 \includegraphics[width=0.45\linewidth]{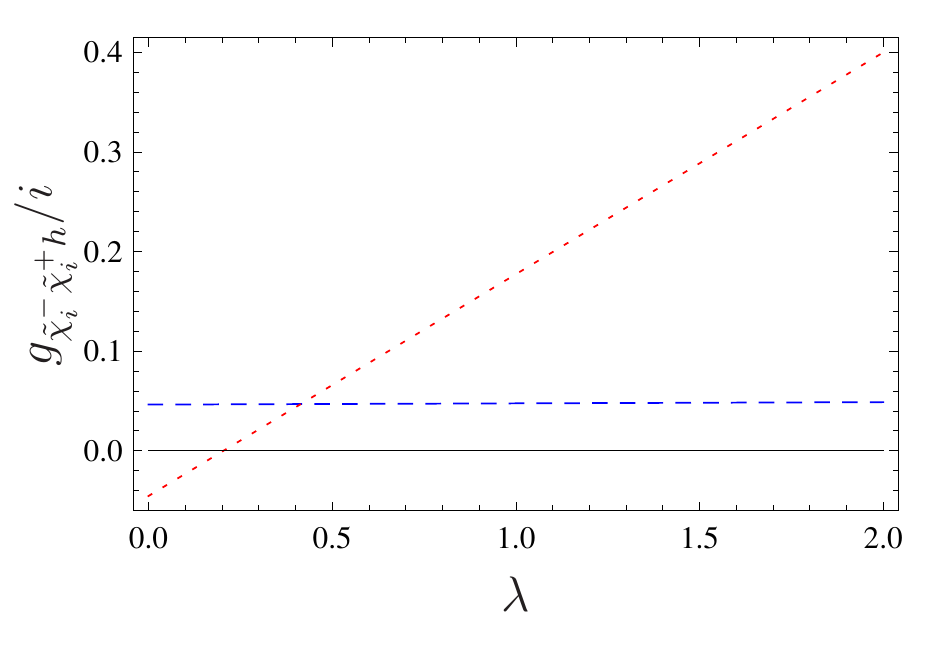} 
\end{minipage}
\caption{The chargino couplings as a function of $\lambda$. The dashed, blue line is for the wino-like chargino and the red, dotted line for the Higgsino-like chargino. Note, to show only the dependence on $\lambda$, we kept the involved rotation matrices in the vertex constant for all values of $\lambda$. The Higgs and chargino mixing matrices have been calculated using the input values $\tan\beta = 1.3$, $\lambda = 0.8$, $\kappa = 0.9$, $\lambda A_\lambda = 23$~GeV, $\kappa A_\kappa = -400$~GeV, $v_s = 265$~GeV, $M_2 = 2$~TeV, and all non-NMSSM parameters have been set to zero. This leads to
$Z^h \simeq (0.61, 0.72, 0.32)$. }
\label{fig:ChaLoop}
\end{figure}
\begin{figure}[hbt]
\begin{minipage}{\linewidth}
 \includegraphics[width=0.45\linewidth]{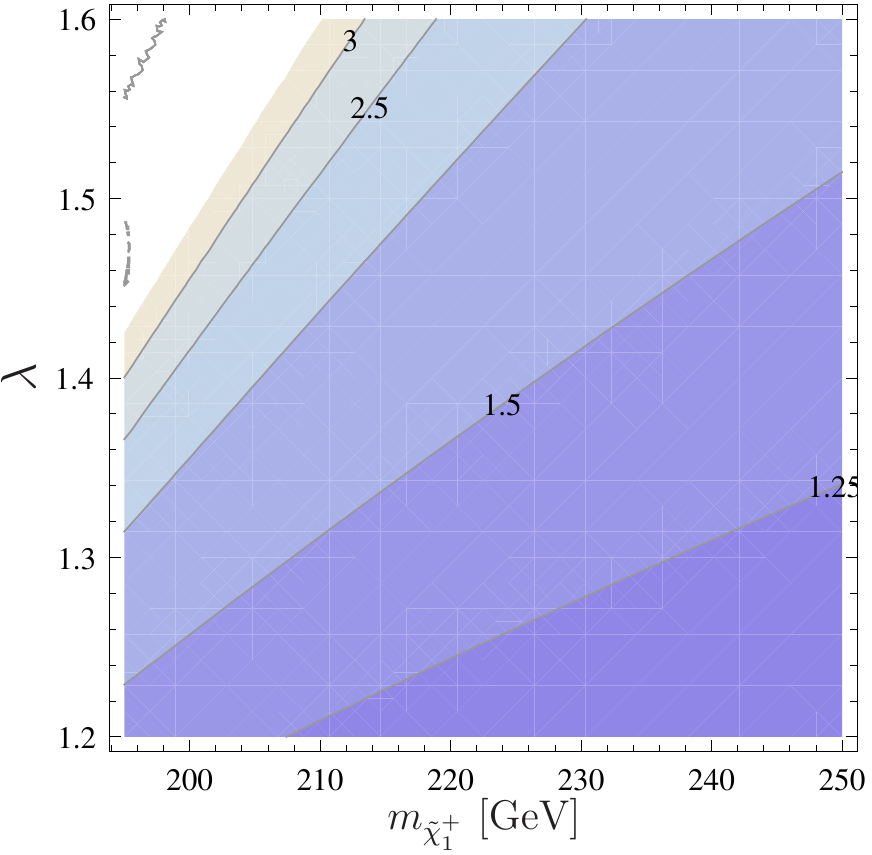} \hfill
\includegraphics[width=0.45\linewidth]{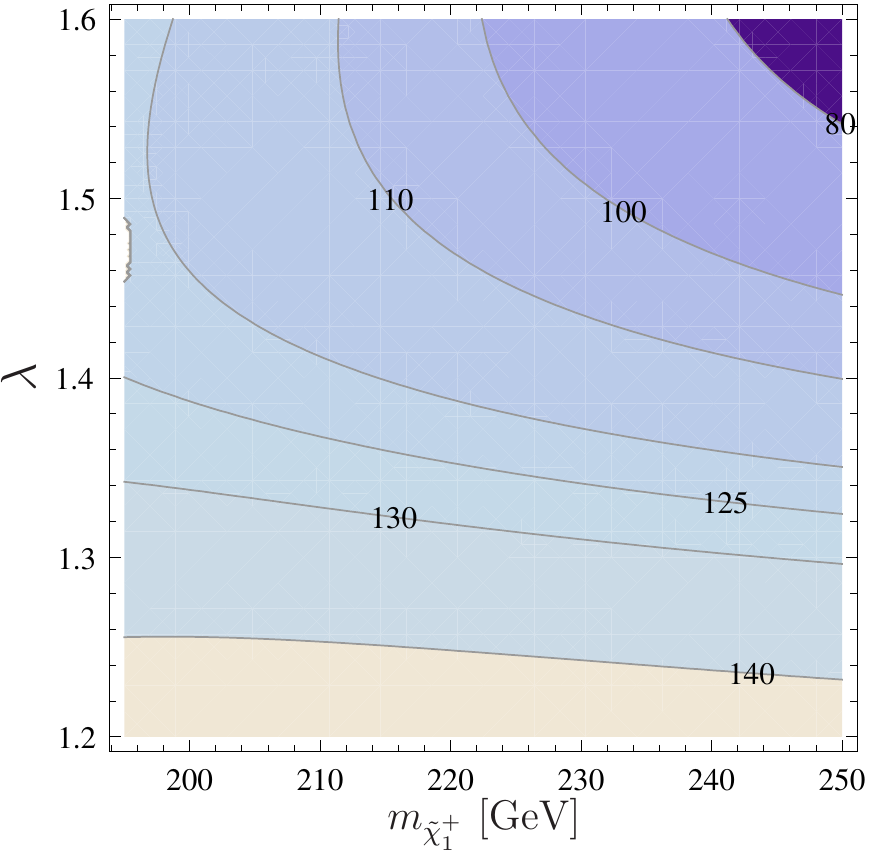}  
\end{minipage}
\caption{Approximate analysis at tree level in the $(m_{\tilde{\chi}^+_1},\lambda)$ plane. Left panel: partial di-photon width from the chargino normalised to the top and $W$ contributions. Right panel: tree-level mass of the light doublet Higgs.  The other parameters are $\tan\beta = 1.2$, $\kappa = 1.1$,  $m_{H^\pm} = 180$~GeV and $\kappa A_{\kappa} = -500$~GeV. }
\label{fig:CharginoTree}
\end{figure}

The chargino - Higgs vertex $\tilde{\chi}^-_i (P_L g^L + P_R g^R) \tilde{\chi}^+_j h$ can be expressed as
\begin{align}
g_{\tilde{\chi}^-_{{{i}}}\tilde{\chi}^+_{{{j}}}h}^L = & \,
-i \frac{1}{\sqrt{2}} \Big(g_2 V^*_{{j} 1} U^*_{{i} 2} Z^h_{ 2}  + V^*_{{j} 2} \Big(g_2 U^*_{{i} 1} Z^h_{ 1}  + \lambda U^*_{{i} 2} Z^h_{ 3} \Big)\Big) \; ,\\
g_{\tilde{\chi}^-_{{{i}}}\tilde{\chi}^+_{{{j}}}h}^R =  &\, (g_{\tilde{\chi}^-_{{{i}}}\tilde{\chi}^+_{{{j}}}h}^L)^* \; .
\end{align}
The unitary matrices which diagonalise the chargino mass matrix can be expressed by two rotation matrices with the angles $\Psi$ and $\Phi$. For the interaction of the light chargino we can write explicitly
\begin{equation}
g_{\tilde{\chi}^-_{{{i}}}\tilde{\chi}^+_{{{j}}}h}^L = -\frac{i}{\sqrt{2}}\Big\{ g_2 (\cos\Psi \sin\Phi Z^h_{ 1} + \sin\Psi \cos\Phi Z^h_{ 2}) + \lambda \sin\Psi \sin\Phi  Z^h_{3}\Big\} \: .
\end{equation}
The first observation is that unlike in the case of the charged Higgs, the couplings to the neutral doublet Higgs components are MSSM like and additional terms appear only 
due to mixing with the singlet state.
To get more insight in this expression we can take the limit $\tan\beta \to 1$ for which $\Phi \sim \Psi$, leading to
\begin{equation}
g_{\tilde{\chi}^-_{{{i}}}\tilde{\chi}^+_{{{j}}}h}^L = -\frac{i}{\sqrt{2}}\left(g_2 \cos\Psi\sin\Psi (Z^h_{1} + Z^h_{2}) +\lambda  \sin^2\Psi  Z^h_{3} \right) \; .
\end{equation} 
There are two limits of interest:
\begin{itemize}
 \item The light chargino is mostly wino-like ($\Psi \to 0$): the coupling to the Higgs is very suppressed.
 \item The light chargino is a Higgsino: ($\Psi \to \pi/2$): the first term vanishes and depending and the sign of $\lambda$ and $Z^h_{3}$ the second term can contribute positively or negatively.
\end{itemize}
The coupling $g_{\tilde{\chi}^-_{{{i}}}\tilde{\chi}^+_{{{j}}}h}$ as a function of $\lambda$ is shown in Figure~\ref{fig:ChaLoop} for the wino and Higgsino limit.
The corresponding enhancement of the partial di-photon width due to the light chargino relative to the SM contributions as well as the tree-level mass of the light Higgs are shown in Figure~\ref{fig:CharginoTree}.  It can be seen that for light charginos and large values of $\lambda$, the chargino loop can be several times larger than the sum of the top and $W$-boson. 

It has recently be pointed out that the chargino loop can also be important in $SU(2)_L$ triplet extensions of the MSSM \cite{Delgado:2012sm}. Similarly to our case, also there large superpotential couplings between the Higgs and the triplets must be present to generate a large enough loop contribution.

\section{Analysis including the Higgs mass and other constraints}
\label{sec:spheno}
After we have discussed the basic principle in the last section at tree-level, we turn now to a full-fledged numerical analysis. Given that we try to explain an enhanced Higgs branching ratio into two photons with rather light charged particles in the loop, we have to be careful not to cause dangerous contributions to precision observables. This leads to additional constraints on the model. The most important observables are briefly discussed in the following:
\paragraph{Anomalous magnetic moment of the muon:} $(g-2)_\mu$ of the muon is almost orthogonal to the di-photon rate in our setup, in contrast to what was argued in the MSSM light stau scenario with slepton mass universality~\cite{Giudice:2012pf}. The reason is that the slepton masses and mixings and correspondingly $g-2$ can be changed without changing the di-photon rate.  While it can happen that there are large contributions from pseudoscalars \cite{Leveille:1977rc,Giudice:2012ms}, these contributions can always be cancelled against the slepton contributions.
\paragraph{$\rho$ - parameter:} the term $\lambda S H_u H_d$ breaks the custodial $SU(2)_L$ present in the MSSM and SM. In general this can cause large contributions for instance to $\delta\rho=1-\rho=\frac{\Pi_{WW}(0)}{m_W^2}-\frac{\Pi_{ZZ}(0)}{m_Z^2}$ where $\Pi_{ZZ}$, $\Pi_{WW}$ are the self-energies of the massive vector bosons. However, the size of the contributions caused by the singlet interaction is much smaller than in the case of a triplet interaction  $\lambda_T H_u T H_d$ and usually save \cite{Nelson:2002ca,Delgado:2012sm}.
\paragraph{Top decays:} A light charged Higgs boson can open new decay channels of the top quark like $t \to H^+ b \to b e^+ \nu$. Recent searches at the LHC put upper bounds on the BR($t \to H^+ b)$ in the range of 2-3\% \cite{Aad:2012tj,CMS:2012cw}.
\paragraph{Radiative $b$ decay:} a very severe constraint comes from $\text{BR}(b \to s \gamma)$, which we will therefore discuss in a bit more detail.\footnote{We would like to thank U.~Haisch for illuminating discussions on this point.} The ratio of SUSY to SM contributions can be written as \cite{Misiak:2004ew,Misiak:2006ab,Misiak:2006zs,Freitas:2008vh}
\begin{equation}
R \equiv \frac{\text{BR}(b \to s \gamma)_\text{SUSY}}{\text{BR}(b \to s \gamma)_\text{SM}} \simeq 1 - 2.55 \Delta C_7 + 1.57 (\Delta C_7)^2 \thickspace ,
\end{equation}
where $\Delta C_7$ are the 
new physics contributions to the Wilson coefficient of the electromagnetic dipole operator (in our case mainly due to the charged Higgs and chargino states). Adding to the  uncertainty of the SM prediction ${\rm Br} (B \to X_s \gamma)_{\rm SM} = (3.15 \pm 0.23) \cdot 10^{-4}$~\cite{Misiak:2006zs,Misiak:2006ab} an intrinsic SUSY error of $0.15$ as well as the error of the experimental world average ${\rm Br} (B \to X_s \gamma)_{\rm exp} = (3.43 \pm 0.22) \cdot 10^{-4}$~\cite{Amhis:2012bh}, leads to the following $95\%$~CL bound 
\begin{equation}
R = [0.87, 1.31] \, .
\end{equation}
In the limit of a pure Higgsino-like chargino and a decoupled wino, these contributions can be approximated by
\begin{eqnarray}
 \Delta C_7^{H^+} &\simeq&  \frac{1}{\tan^{2}\beta} f_1(m^2_t/m^2_{H^+}) + f_2(m^2_t/m^2_{H^+}) \;,\\
 \Delta C_7^{\tilde{\chi}^+} &\simeq& - \frac{m_t^2}{m_{\tilde{t}}^2} \frac{M_2^2}{\mu_\text{eff}^2} f_1(m_{\tilde{t}}^2/M_2^2) - \mu X_t \tan\beta \frac{m_t^2}{m_{\tilde{t}}^4} \frac{M_2^2}{\mu_\text{eff}^2} f_3(m_{\tilde{t}}^2/M_2^2) \; ,
\end{eqnarray}
with $X_t = A_t - \mu_\text{eff}/\tan\beta$ and loop functions \cite{Barbieri:1993av,Uli}
\begin{eqnarray}
 f_1(x)&=&\frac{7x-5x^2-8x^3}{72(x-1)^3}+\frac{3x^3-2x^2}{12(x-1)^4}\ln x\; ,\nonumber \\
 f_2(x)&=&\frac{3x-5x^2}{12(x-1)^2}+\frac{3x^2-2x}{6(x-1)^3} \ln x\; ,\nonumber \\
 f_3(x)&=&\frac{7x^2-13x^3}{12(x-1)^3}+\frac{3x^4+2x^3-2x^2}{6(x-1)^4}\ln x \; .
\end{eqnarray}

\begin{figure}[hbt]
\begin{minipage}{\linewidth}
 \includegraphics[width=0.45\linewidth]{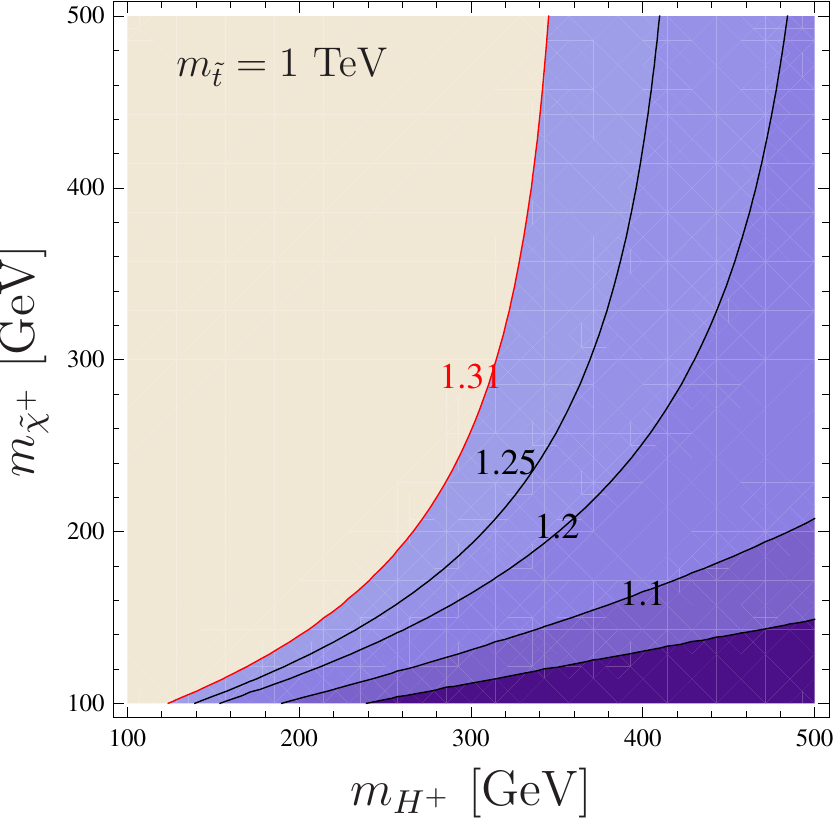} \hfill
\includegraphics[width=0.45\linewidth]{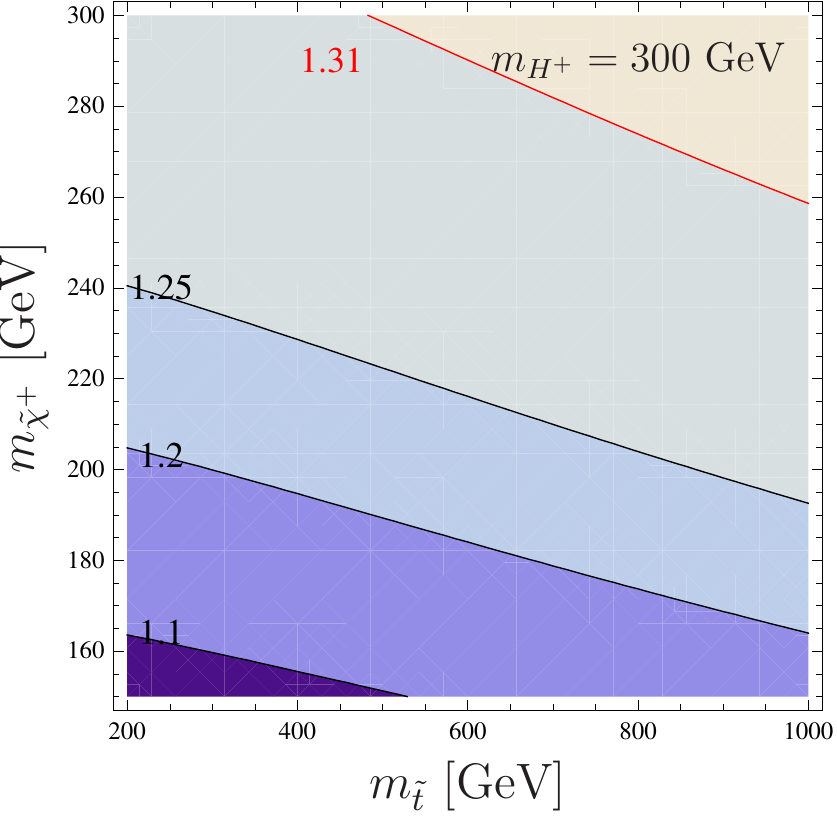}  
\end{minipage}
\caption{The ratio $R$ as a function of the chargino mass and the charged Higgs mass (left) and stop mass (right). 
In both plots $X_t = 0$, $M_2 = 2\tev$ and $\tan\beta=1$ have been used.  }
\label{fig:DeltaC7}
\end{figure}
While the charged Higgs loop depends only on the mass of the Higgs (and the top quark), the chargino loop depends on the stop sector as well as on the wino mass $M_2$ in addition. In order to evade the experimental bound on $R$ either the charged Higgs has
to be sufficiently heavy, or there has to be a cancellation between the contributions from the charged Higgs and the chargino.
To maximize the chargino loop, $M_2$ has to be large and light stops are needed. Figure~\ref{fig:DeltaC7} shows the value of $R$ as a function
of the chargino and charged Higgs as well as the stop mass, to illustrate the contraints from $b \rightarrow s \gamma$.
It can be seen that $m_{\tilde{\chi}^+}$ has to be smaller than $m_{{H}^+}$ unless $m_{{H}^+} \gtrsim 350 \gev$.
Typically the main contribution to the di-photon rate will therefore originate from the charginos.

To calculate the Higgs mass, the diphoton rate as well as the precision observables, we use the \SPheno version \cite{Porod:2003um,Porod:2011nf} for the GNMSSM created by \SARAH \cite{Staub:2008uz,Staub:2009bi,Staub:2010jh,Staub:2012pb} which has been presented in \cite{Ross:2012nr}. This version performs a complete one-loop calculation of all SUSY and Higgs masses \cite{Pierce:1996zz,Staub:2010ty} and includes the dominant two-loop corrections for the scalar Higgs masses \cite{Dedes:2003km,Dedes:2002dy,Brignole:2002bz,Brignole:2001jy}. In addition, it calculates observables like $b \to s \gamma$ and $g-2$ for the given model with the same precision as in the MSSM including all possible new contributions. 
We performed an analysis in terms of the low-energy variables and fixed the slepton soft-breaking mass squareds of the first two generations to $5\cdot 10^5~\text{GeV}^2$ and those of the third generation to $10^5~\text{GeV}^2$. In the squark sector we used $5\cdot 10^6~\text{GeV}^2$ for the first two generations, and $10^6~\text{GeV}^2$ for the third one. Finally, all MSSM A-terms except $A_t$ were put to zero. 

In our scan we searched for parameter values which realise a significant enhancement of $pp\rightarrow h \rightarrow \gamma\gamma$ due to light charginos in the loop, taking into account both Higgs production and branching ratios.

\begin{figure}[hbt]
\begin{minipage}{\linewidth}
\includegraphics[width=0.32\linewidth]{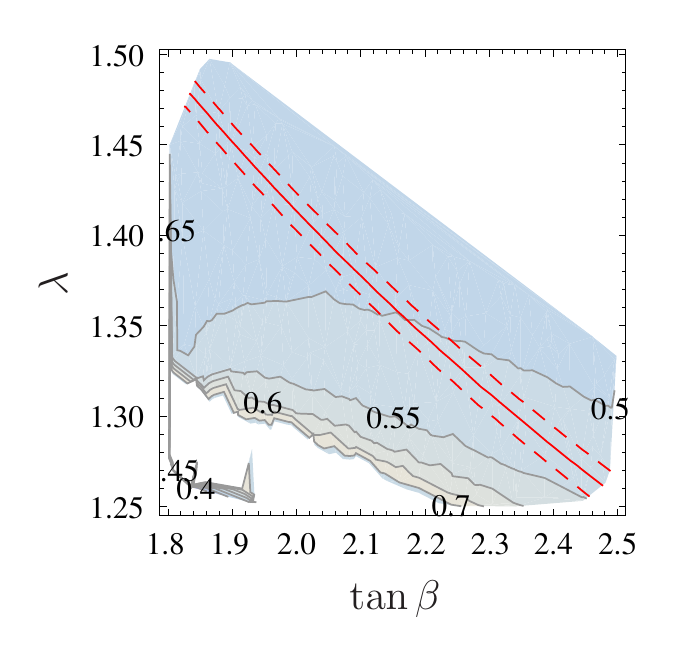}  \hfill
\includegraphics[width=0.32\linewidth]{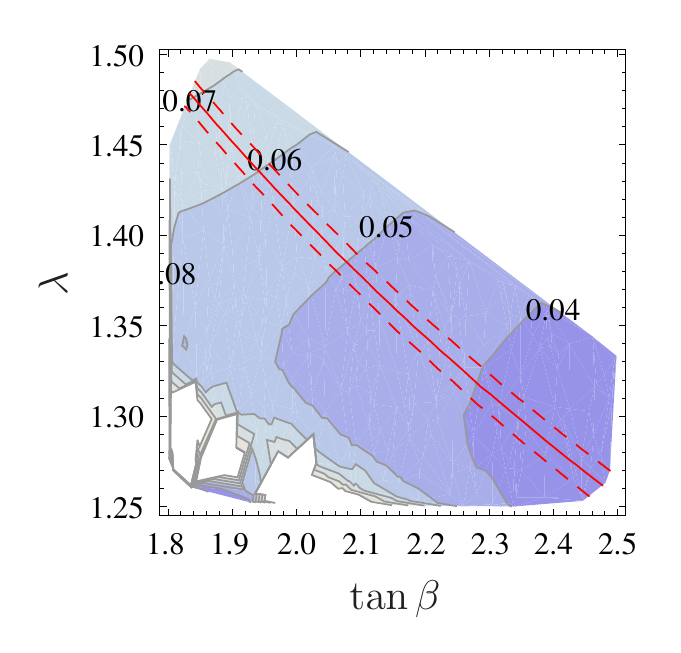}  \hfill
\includegraphics[width=0.32\linewidth]{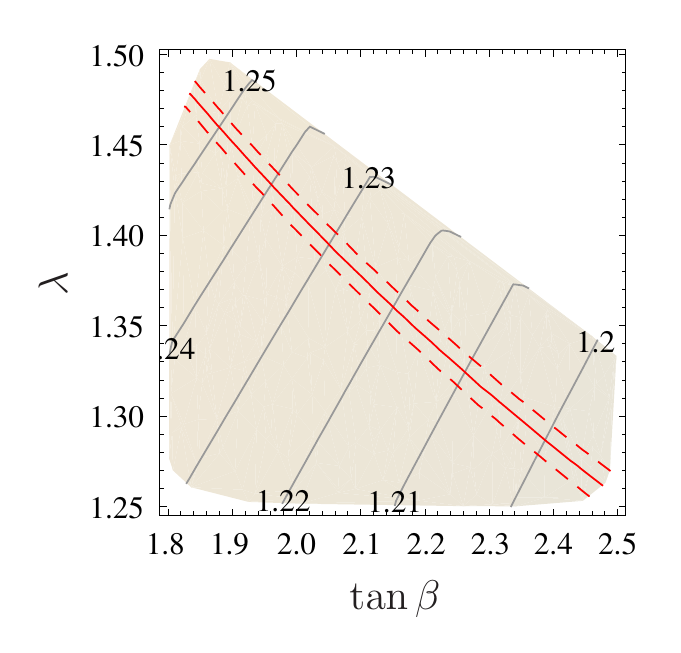} 
\end{minipage}
\caption{Full one-loop evaluation of the ratio of the partial di-photon width from the charginos (left) and charged Higgs (middle) normalised to the standard model like contribution in the $(\tan\beta,\lambda)$-plane. The right figure shows the corresponding value of $R$ as calculated by {\tt SPheno}. The red line corresponds to a light Higgs mass of 125 GeV with $\pm 3$ GeV indicated by the dashed lines.  In both plots the other parameters were fixed to the values given in Eq.~(\ref{eq:inputvalues}). }
\label{fig:BPpartial}
\end{figure}

As an example we present a benchmark point with the following input parameters,
\begin{equation}
\label{eq:inputvalues}
\begin{array}{c}
\kappa = 0.2 \,, \,\, \lambda A_\lambda = 330~\text{GeV} \,, \,\, \kappa A_\kappa = 14~\text{GeV} \,, \,\, v_s = 166~\text{GeV} \,, \\
 \mu = -21~\text{GeV}\,, \,\, \mu_S = 68~\text{GeV}\,, \,\, b\mu = 3151~\text{GeV}^2\,, \,\, A_t Y_t = -432~\text{GeV} 
\end{array}
\end{equation}
and gaugino masses of $M_2 = 2 \cdot M_1 = 2$~TeV, $M_3 = 1.4$~TeV. Note that $M_3$ has just been chosen to fulfil the current constraint from direct searches, but the concrete value has only a small impact on our discussion. A point with another value of $M_3$ fulfilling $M_3 > M_2 > M_1$ would have worked equally well. 
\begin{figure}[hbt]
\begin{minipage}{\linewidth}
 \includegraphics[width=0.45\linewidth]{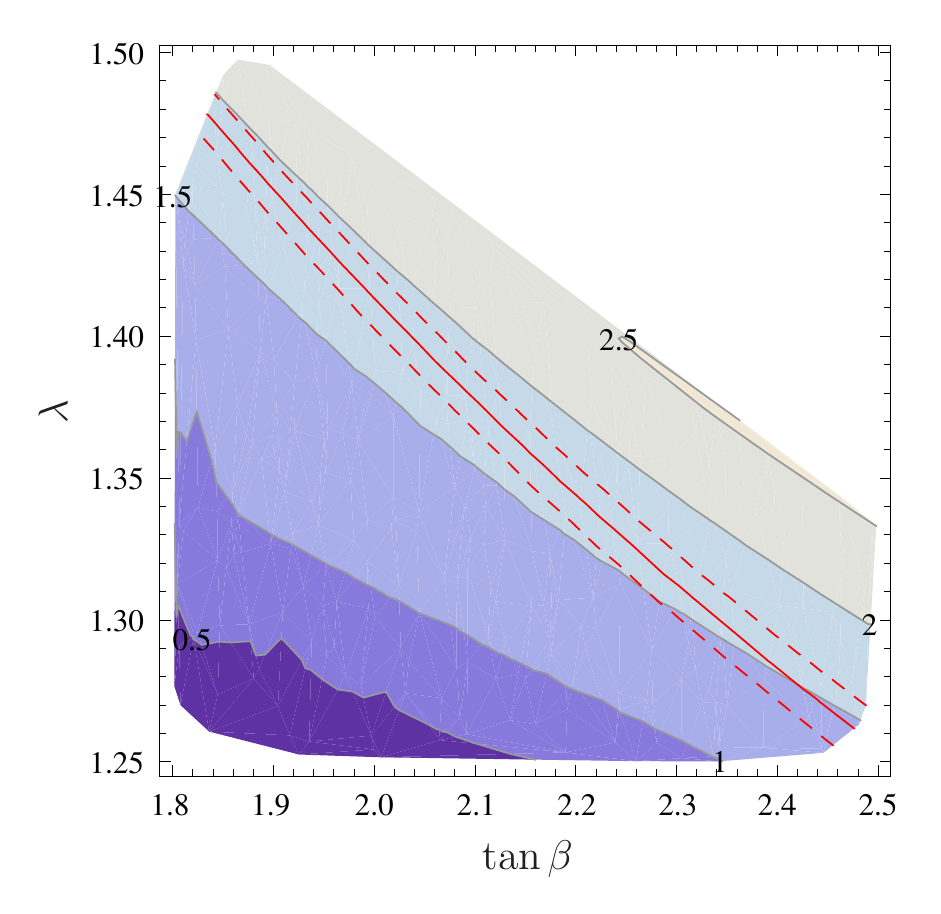} \hfill
\includegraphics[width=0.45\linewidth]{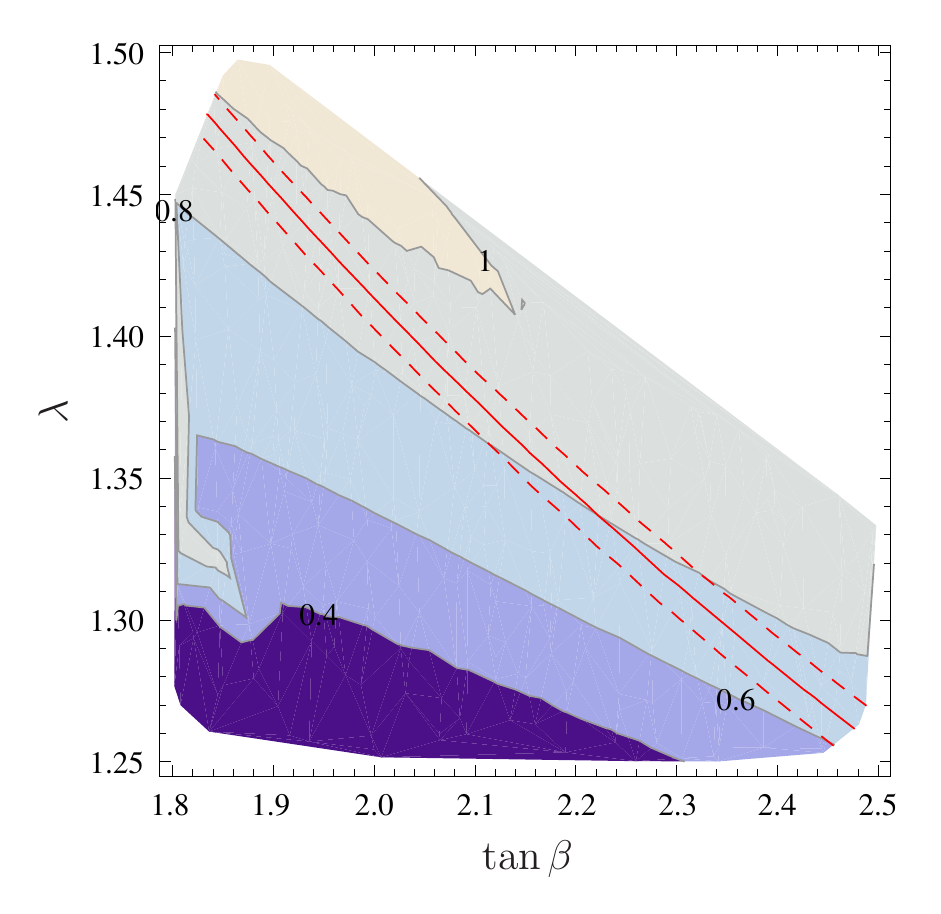} 
\end{minipage}
\caption{Full one-loop Higgs production cross section times branching into $\gamma \gamma$ (left) and $WW^*$ (right) in the $(\tan\beta,\lambda)$-plane. The red line corresponds to a light Higgs mass of 125 GeV with $\pm 3$ GeV indicated by the dashed lines.         It can be seen that a significant enhancement in the $\gamma\gamma$ channel is possible with the $WW^*$ channel SM-like. The other parameters are given in the text.}
\label{fig:BPsigmaBR}
\end{figure}
Let us start with looking at the relative contributions of the chargino: the results for the partial di-photon width due to the chargino loop normalised to the standard model contributions is shown in Figure~\ref{fig:BPpartial}. 
It can be seen that for e.g.\ $\tan\beta \sim 2.3$ and $\lambda \sim 1.3$ the chargino loop adds more than 50\% to the SM contribution. In Figure~\ref{fig:BPsigmaBR} we show the corresponding Higgs production cross section (from gluon and vector boson fusion) times branching ratio for the channels $\gamma \gamma$ and $WW^*$, normalised to the SM.
We see that for $\tan\beta \sim 2.3$ and $\lambda \sim 1.3$ we obtain an enhancement in the diphoton rate of about 50\%
while the corresponding ratio for  $W W^*$ production is roughly 1. Also the Higgs mass is in the preferred range of 122-128~GeV as indicated  by the dashed red lines. For this point the charged Higgs mass is 
$m_{H^+} = 323 \gev$ while the chargino mass is                
$m_{\tilde{\chi}_1^+} = 128 \gev$.

\section{Summary and Conclusions}
\label{sec:conclusions}

In this article we have shown that loops involving light charged Higgs as well as chargino states, while negligible within the framework of the MSSM, can significantly contribute to
the partial Higgs di-photon decay width in MSSM singlet extensions. Such an enhancement typically requires rather large values of the doublet-singlet coupling $\lambda$, corresponding to the well-known
$\lambda$SUSY scenario. What is particularly interesting about our findings is that large $\lambda$ is preferred independently from the recent experimental indications in the Higgs sector,
as it enhances the tree-level Higgs mass and alleviates the electroweak fine tuning.

\section*{Acknowledgements}
We would like to thank Graham G.~Ross, Mark Goodsell and Ulrich Haisch for useful discussions. The research presented here was partially supported by the EU ITN grant UNILHC 237920 (Unification in the LHC era) and the ERC Advanced Grant BSMOXFORD 228169. 

\begin{appendix}
\section{Mass matrices}
\label{app:massmatrices}
After electroweak symmetry breaking, the Higgs fields are decomposed as follows:
\begin{align} 
H_d^0 = & \, \frac{1}{\sqrt{2}} \left( h_{d}  +  v_d  + i \sigma_{d} \right) \\ 
H_u^0 = & \, \frac{1}{\sqrt{2}} \left( h_{u}  +  v_u  + i  \sigma_{u} \right) \\ 
S = & \, \frac{1}{\sqrt{2}}  \left( s  + v_s  + i \sigma_s  \right)
\end{align} 

\begin{itemize}
\item {\bf Mass matrix for CP even Higgs} \\ 
The matrix for the CP even Higgs can be written in the basis \( \left(h_{d}, h_{u}, s\right)^T\) as
\begin{equation} 
m^2_{h} = \left( 
\begin{array}{ccc}
m_{dd} &m_{du} &m_{ds}\\ 
m_{du} &m_{uu} &m_{us}\\ 
m_{ds} &m_{us} &m_{ss}\end{array} 
\right) 
\end{equation} 
with
\begin{align} 
m_{dd} &= \frac{1}{8} \Big(8 m_{H_d}^2 +3 g_{1}^{2} v_{d}^{2} +3 g_{2}^{2} v_{d}^{2} - g_{1}^{2} v_{u}^{2} - g_{2}^{2} v_{u}^{2} +4 \Big(\sqrt{2} v_s \mu  + v_{s}^{2} \lambda  + v_{u}^{2} \lambda \Big)\lambda^* \nonumber \\ 
 &+4 \Big(2 \mu  + \sqrt{2} v_s \lambda \Big)\mu^* \Big)\\ 
m_{du} &= \frac{1}{4} \Big(- g_{1}^{2} v_d v_u - g_{2}^{2} v_d v_u +4 v_d v_u |\lambda|^2 -2 \lambda \xi^* - \sqrt{2} v_s \lambda \mu_s^* - v_{s}^{2} \lambda \kappa^* -2 \xi \lambda^* \nonumber \\ 
 &- \sqrt{2} \mu_s v_s \lambda^* - v_{s}^{2} \kappa \lambda^* - \sqrt{2} v_s (\lambda A_\lambda)^* -4 {\Re\Big(b\mu\Big)} - \sqrt{2} v_s \lambda A_\lambda \Big)\\ 
m_{uu} &= \frac{1}{8} \Big(8 m_{H_u}^2 - g_{1}^{2} v_{d}^{2} - g_{2}^{2} v_{d}^{2} +3 g_{1}^{2} v_{u}^{2} +3 g_{2}^{2} v_{u}^{2} +4 \Big(v_{d}^{2} \lambda  + v_s \Big(\sqrt{2} \mu  + v_s \lambda \Big)\Big)\lambda^* \nonumber \\ 
 &+4 \Big(2 \mu  + \sqrt{2} v_s \lambda \Big)\mu^* \Big)\\ 
m_{ds} &= \frac{1}{4} \Big(4 v_d v_s |\lambda|^2 - \sqrt{2} v_u \lambda \mu_s^* -2 v_s v_u \lambda \kappa^* - \sqrt{2} \mu_s v_u \lambda^* -2 v_s v_u \kappa \lambda^* +2 \sqrt{2} v_d \mu \lambda^* \nonumber \\ 
 &+2 \sqrt{2} v_d \lambda \mu^* -2 \sqrt{2} v_u {\Re\Big(\lambda A_\lambda\Big)} \Big)\\ 
m_{us} &= \frac{1}{4} \Big(4 v_s v_u |\lambda|^2 - \sqrt{2} v_d \lambda \mu_s^* -2 v_d v_s \lambda \kappa^* - \sqrt{2} \mu_s v_d \lambda^* -2 v_d v_s \kappa \lambda^* +2 \sqrt{2} v_u \mu \lambda^* \nonumber \\ 
 &+2 \sqrt{2} v_u \lambda \mu^* -2 \sqrt{2} v_d {\Re\Big(\lambda A_\lambda\Big)} \Big)\\ 
m_{ss} &= \frac{1}{2} \Big(2 m_S^2 +6 v_{s}^{2} |\kappa|^2 +v_{d}^{2} |\lambda|^2 +v_{u}^{2} |\lambda|^2 +2 \kappa \xi^* +\Big(2 \mu_s  + 3 \sqrt{2} v_s \kappa \Big)\mu_s^* \nonumber \\ 
 & +2 \xi \kappa^* +3 \sqrt{2} \mu_s v_s \kappa^* - v_d v_u \lambda \kappa^* - v_d v_u \kappa \lambda^* +\sqrt{2} v_s (\kappa A_\kappa)^* +2 {\Re\Big(b_s\Big)} +\sqrt{2} v_s (\kappa A_\kappa) \Big)
\end{align} 
This matrix is diagonalized by an unitary matrix \(Z^H\) as
\begin{equation} 
Z^H m^2_{h} Z^{H,\dagger} = m^{2}_{h,diag} 
\end{equation} 
\item {\bf Mass matrix for CP odd Higgs}\\
The matrix for the CP odd Higgs reads in the basis \( \left(\sigma_{d}, \sigma_{u}, \sigma_s\right)^T\) 
\begin{equation} 
m^2_{A^0} = \left(
\begin{array}{ccc}
m_{dd} &m_{du} &m_{ds}\\ 
m_{du} &m_{uu} &m_{us}\\ 
m_{ds} &m_{us} &m_{ss}\end{array} 
\right) 
\end{equation} 
with
\begin{align} 
m_{dd} &= \frac{1}{8} \Big(8 m_{H_d}^2 +g_{1}^{2} v_{d}^{2} +g_{2}^{2} v_{d}^{2} - g_{1}^{2} v_{u}^{2} - g_{2}^{2} v_{u}^{2} +4 \Big(\sqrt{2} v_s \mu  + v_{s}^{2} \lambda  + v_{u}^{2} \lambda \Big)\lambda^* \nonumber \\ 
 &+4 \Big(2 \mu  + \sqrt{2} v_s \lambda \Big)\mu^* \Big)\\ 
m_{du} &= \frac{1}{4} \Big(2 \xi \lambda^*  + 2 \lambda \xi^*  + 4 {\Re\Big(b\mu\Big)}  + \sqrt{2} \mu_s v_s \lambda^*  + \sqrt{2} v_s \lambda \mu_s^* \nonumber \\ &  + \sqrt{2} v_s \lambda A_\lambda  + \sqrt{2} v_s (\lambda A_\lambda)^*  + v_{s}^{2} \kappa \lambda^*  + v_{s}^{2} \lambda \kappa^* \Big)\\ 
m_{uu} &= \frac{1}{8} \Big(8 m_{H_u}^2 - g_{1}^{2} v_{d}^{2} - g_{2}^{2} v_{d}^{2} +g_{1}^{2} v_{u}^{2} +g_{2}^{2} v_{u}^{2} +4 \Big(v_{d}^{2} \lambda  + v_s \Big(\sqrt{2} \mu  + v_s \lambda \Big)\Big)\lambda^* \nonumber \\ 
 &+4 \Big(2 \mu  + \sqrt{2} v_s \lambda \Big)\mu^* \Big)\\ 
m_{ds} &= -\frac{1}{4} v_u \Big(-2 \sqrt{2} {\Re\Big(\lambda A_\lambda\Big)}  + 2 v_s \kappa \lambda^*  + 2 v_s \lambda \kappa^*  + \sqrt{2} \lambda \mu_s^*  + \sqrt{2} \mu_s \lambda^* \Big)\\ 
m_{us} &= -\frac{1}{4} v_d \Big(-2 \sqrt{2} {\Re\Big(\lambda A_\lambda\Big)}  + 2 v_s \kappa \lambda^*  + 2 v_s \lambda \kappa^*  + \sqrt{2} \lambda \mu_s^*  + \sqrt{2} \mu_s \lambda^* \Big)\\ 
m_{ss} &= \frac{1}{2} \Big(2 m_S^2 +2 v_{s}^{2} |\kappa|^2 +v_{d}^{2} |\lambda|^2 +v_{u}^{2} |\lambda|^2 -2 \kappa \xi^* +\Big(2 \mu_s  + \sqrt{2} v_s \kappa \Big)\mu_s^* \\ 
 & -2 \xi \kappa^* +\sqrt{2} \mu_s v_s \kappa^* \nonumber +v_d v_u \lambda \kappa^* +v_d v_u \kappa \lambda^* - \sqrt{2} v_s (\kappa A_\kappa)^* -2 {\Re\Big(b_s\Big)} - \sqrt{2} v_s (\kappa A_\kappa) \Big)
\end{align} 
This matrix is diagonalized by \(Z^A\): 
\begin{equation} 
Z^A m^2_{A^0} Z^{A,\dagger} = m^{2}_{A^0,diag} 
\end{equation} 

\item {\bf Mass matrix for charged Higgs} \\
The charged Higgs mass matrix reads in the basis \( \left(H_d^-, H_u^{+,*}\right)^T\) 
\begin{equation} 
m^2_{H^-} = \left( 
\begin{array}{cc}
m_{dd} &m^*_{du}\\ 
m_{du} &m_{uu}\end{array} 
\right) 
\end{equation} 
with
\begin{align} 
m_{dd} &= \frac{1}{8} \Big(4 \Big(2 \mu  + \sqrt{2} v_s \lambda \Big)\mu^*  + 4 v_s \Big(\sqrt{2} \mu  + v_s \lambda \Big)\lambda^*  + 8 m_{H_d}^2  + (g_{1}^{2} + g_{2}^{2})( v_{d}^{2}  + v_{u}^{2}) \Big)\\ 
m_{du} &= \frac{1}{4} \Big(2 \Big(2 \xi  + \sqrt{2} \mu_s v_s  - v_d v_u \lambda  + v_{s}^{2} \kappa \Big)\lambda^*  + 2 \sqrt{2} v_s (\lambda A_\lambda)^*  + 4 b\mu^*  + g_{2}^{2} v_d v_u \Big)\\ 
m_{uu} &= \frac{1}{8} \Big(4 \Big(2 \mu  + \sqrt{2} v_s \lambda \Big)\mu^*  + 4 v_s \Big(\sqrt{2} \mu  + v_s \lambda \Big)\lambda^*  + 8 m_{H_u}^2  - (g_{1}^{2} + g_{2}^{2})( v_{d}^{2}  + v_{u}^{2}) \Big)
\end{align} 
This matrix is diagonalized by \(Z^+\): 
\begin{equation} 
Z^+ m^2_{H^-} Z^{+,\dagger} = m^{2}_{H^-,diag} 
\end{equation} 

\item {\bf Mass matrix for Neutralinos} \\
The neutralino mass matrix reads in the basis: \( \left(\lambda_{\tilde{B}}, \tilde{W}^0, \tilde{H}_d^0, \tilde{H}_u^0, \tilde{S}\right)^T \) 
 
\begin{equation} 
m_{\tilde{\chi}^0} = \left( 
\begin{array}{ccccc}
M_1 &0 &-\frac{1}{2} g_1 v_d  &\frac{1}{2} g_1 v_u  &0\\ 
0 &M_2 &\frac{1}{2} g_2 v_d  &-\frac{1}{2} g_2 v_u  &0\\ 
-\frac{1}{2} g_1 v_d  &\frac{1}{2} g_2 v_d  &0 &- \frac{1}{\sqrt{2}} v_s \lambda  - \mu  &- \frac{1}{\sqrt{2}} v_u \lambda \\ 
\frac{1}{2} g_1 v_u  &-\frac{1}{2} g_2 v_u  &- \frac{1}{\sqrt{2}} v_s \lambda  - \mu  &0 &- \frac{1}{\sqrt{2}} v_d \lambda \\ 
0 &0 &- \frac{1}{\sqrt{2}} v_u \lambda  &- \frac{1}{\sqrt{2}} v_d \lambda  &\sqrt{2} v_s \kappa  + \mu_s\end{array} 
\right) 
\end{equation} 
This matrix is diagonalized by \(N\): 
\begin{equation} 
N m_{\tilde{\chi}^0} N^{\dagger} = m^{dia}_{\tilde{\chi}^0} 
\end{equation} 

\item {\bf Mass matrix for Charginos} \\
The chargino mass matrix reads in the basis \( \left(\tilde{W}^-, \tilde{H}_d^-\right)^T / \left(\tilde{W}^+, \tilde{H}_u^+\right) \) 
 
\begin{equation} 
m_{\tilde{\chi}^-} = \left( 
\begin{array}{cc}
M_2 &\frac{1}{\sqrt{2}} g_2 v_u \\ 
\frac{1}{\sqrt{2}} g_2 v_d  &\frac{1}{\sqrt{2}} v_s \lambda  + \mu\end{array} 
\right) 
\end{equation} 
To diagoanlize this non-symmetric matrix two unitary matrices \(U\) and \(V\)  are needed:
\begin{equation} 
U^* m_{\tilde{\chi}^-} V^{\dagger} = m^{dia}_{\tilde{\chi}^-} 
\end{equation} 
\end{itemize}  
\end{appendix}

\bibliography{Gamma}
\bibliographystyle{ArXiv}

\end{document}